\documentstyle[aps,prb,subfigure,epsfig,amsmath,amssymb]{revtex}
\begin{document}
\draft

\newcommand {\be}{\begin{equation}}
\newcommand {\ee}{\end{equation}}
\newcommand {\bea}{\begin{eqnarray}}
\newcommand {\eea}{\end{eqnarray}}
\newcommand {\refeq}[1] {(\ref{#1})}
\newcommand {\vett}[1] {{\mathbf{#1}}}
\newcommand {\rvec}{{\bf r}}

\title{Quasicondensate and superfluid fraction in the 2D
charged-boson gas at finite temperature}
\author{E. Strepparola, A. Minguzzi and M. P. Tosi}
\address{Istituto Nazionale di Fisica della Materia and
Classe di Scienze, Scuola Normale Superiore, Piazza dei Cavalieri 7,
56126 Pisa, Italy}

\maketitle                
\begin{abstract}
The Bogoliubov - de Gennes equations are
solved for the Coulomb Bose gas describing a fluid of charged bosons
at finite temperature. The approach is applicable in the weak coupling
regime and the extent of its quantitative usefulness is tested in the
three-dimensional fluid, for which diffusion Monte Carlo data are
available on the condensate fraction at zero temperature. The one-body
density matrix is then evaluated by the same approach for the
two-dimensional fluid with $e^2/r$ interactions, to demonstrate the
presence of a quasi-condensate from its power-law decay with
increasing distance and to evaluate the superfluid fraction as a
function of temperature at weak coupling.
\end{abstract}

\pacs{PACS numbers: 05.30.Jp, Boson systems}

\section{Introduction}
The fluid of
pointlike spinless charged bosons embedded in a uniform neutralizing
background has attracted attention in the literature mainly as a model
in quantum statistical mechanics, which is complementary to the
physically more relevant fermionic gas of electrons\cite{uno}. It was
proposed by Schafroth\cite{due} as a model for superconductors prior to the
BCS theory and has received renewed interest after the discovery of
ceramic superconductors\cite{tre}. In some viewpoints a Bose-Einstein
condensate of tightly bound pairs of small polarons could be a
relevant model for high-$T_c$ superconductivity in the layered
cuprates\cite{quattro}.  The model also has some astrophysical
relevance in describing 
pressure-ionized helium in cold stellar matter\cite{cinque,sei,sette}
and the fusion of 
$\alpha$-particles inside a dense helium plasma\cite{otto,nove}.

A number of theoretical and computational studies have been addressed
to the three-dimensional charged-boson fluid (3D-CBF) at zero
temperature. The properties of main interest for the theory have been
the ground-state energy and structure and the static and dynamic
dielectric response. The early theoretical work was concerned with
evaluating the ground-state energy and the elementary excitations in
the weak coupling (high density) limit\cite{dieci,undici,dodici}. Both
variational 
calculations based on Jastrow-Feenberg wave
functions\cite{tredici,quattordici,quindici,sedici,diciassette,diciotto}
and 
self-consistent treatments of
correlations\cite{diciannove,venti,ventuno} have subsequently 
been used to evaluate the intermediate and strong coupling
regime. Quantal Monte Carlo studies of the
3D-CBF\cite{sedici,ventidue,ventitre,ventiquattro,venticinque} have
covered the 
whole range of coupling strength up to the regime of Wigner
crystallization driven by the Coulomb repulsions.
 Extensive data on the condensate fraction and the momentum
distribution in dependence of the coupling strength have become
available through the diffusion Monte Carlo (DMC) work of Moroni {\it et
al.}\cite{venticinque}.

The properties of the 2D-CBF at zero temperature have also been
investigated with both $e^2/r$ and $\ln(r)$ interactions. In the latter case
Magro and Ceperley\cite{ventisei} have shown, using a sum rule argument from
Pitaevskii and Stringari\cite{ventisette}, that the presence of the
long-wavelength plasmon in the excitation spectrum rules out the
existence of a condensate even at zero temperature. They further
showed from a DMC study that the noncondensed fluid exhibits a
power-law decay of the one-body density matrix and that above a
threshold density the momentum distribution diverges at low momenta,
although no condensate forms.

In the 2D-CBF with $e^2/r$ interactions
(henceforth referred to as quasi-2D or Q2D-CBF) a condensate is
present in the ground state, but its density vanishes at finite
temperature. This behaviour parallels that of the neutral 2D Bose
gas\cite{ventotto,ventinove}.  The theory of correlations in the latter system has been
developed by Kagan, Svistunov and Shlyapnikov\cite{trenta} (see also Kagan
{\it et
al.}\cite{trentuno}). As its temperature is lowered at constant density (or as
its density is increased at constant temperature) across the
Kosterlitz-Thouless transition, a weakly interacting gas enters the
superfluid regime in which the phase correlation length $R_c$ is much
larger than the density correlation length $r_c$. In this situation the
one-body density matrix $\rho(r)$ decays asymptotically to zero with
an 
inverse-power law rather than
exponentially\cite{trentadue,trentatre}. The idea of a 
"quasicondensate" emerges from the behaviour of $\rho(r)$ at intermediate
distances $r_c \ll r \ll R_c$. The local properties of the
quasicondensate are the same as those of a genuine condensate, so that
it turns to the latter as the phase correlation length
starts to exceed the dimensions of the sample. 

The main purpose of the present work is to study this behaviour in the
Q2D-CBF, using the Bogoliubov approach to describe the charged fluid
both in the ground state and at finite temperature in the weak
coupling regime (corresponding in this case to high density). The
Bogoliubov - de Gennes equations are presented for convenience in
Sect.~2
and are first solved for the 3D-CBF in Sect.~3, where the approach is
quantitatively tested at $T = 0$ by comparing its results for the
condensate fraction with the available DMC data. Section 4 reports our
main results regarding the quasicondensate and the superfluid fraction
in the Q2D-CBF. Finally, Sect.~5 gives a brief summary and our
conclusions.

\section{Bogoliubov approach to a charged-boson fluid}
The fluid of charged bosons on a uniform neutralizing background is
described by the Hamiltonian
\begin{equation}
H=\int d\vett r\, \psi^{\dagger}(\vett
r)\left(-\frac{\nabla^2}{2m}-\mu \right)\psi(\vett r) +\frac{1}{2}
\int d\vett r\,\int d\vett r'\, \psi^{\dagger}(\vett 
r)\psi^{\dagger}(\vett r')V(|\vett r-\vett r'|)\psi(\vett r')\psi(\vett r)\;,
\end{equation}
where $V(r)=e^2/r$, $\psi(\vett r)$ is
the field operator and $\mu$ the chemical potential. The role of the
background is to set to zero the average potential felt by each
particle. The coupling strength is measured by the dimensionless
parameter $r_s$, defined by $r_sa_B=(4\pi n/3)^{-1/3}$ in 3D and by
$r_s a_B=(\pi n)^{-1/2}$ in 2D with $a_B$ the Bohr
radius and $n$ the mean particle density.       

We consider first the
fluid at T = 0. The Bogoliubov transformation\cite{trentaquattro} (for recent
presentations see \cite{trentacinque,trentasei}) introduces a
macroscopic order parameter 
$\psi_0$ by writing the field operator as $\psi(\vett r)=\psi_0+\tilde
\psi(\vett r)$. The operator $\tilde \psi(\vett r)$ describes the
gas of Bose particles promoted out of the condensate. This gas is
treated in the Hartree approximation, assuming that the contribution
from terms nonlinear in $\tilde \psi(\vett r)$ is small.

One finds $\mu = 0$ and, with the
linear transformation
\begin{equation}
\tilde \psi(\vett r,t)=\sum_\nu \left[ u_\nu(\vett r,t)
a_\nu+v_\nu^*(\vett r,t) a_\nu^\dagger\right]
\end{equation}
for the Heisenberg field
operator, one has to solve the coupled linear equations
\begin{equation}
i \frac{\partial u_\nu(\vett r,t)}{\partial t}=-\frac{1}{2m} \nabla^2
u_\nu(\vett r,t) +n_0 \int  d\vett r'\, V(|\vett r-\vett r'|)
[u_\nu(\vett r,t)+u_\nu(\vett r',t)+v_\nu(\vett r',t)] 
\end{equation}
and
\begin{equation}
-i \frac{\partial v_\nu(\vett r,t)}{\partial t}=-\frac{1}{2m} \nabla^2
v_\nu(\vett r,t) +n_0 \int  d\vett r'\, V(|\vett r-\vett r'|)
[v_\nu(\vett r,t)+v_\nu(\vett r',t)+u_\nu(\vett r',t)] \;.
\end{equation}
Here, $n_0=\psi_0^2$ is the uniform condensate
density. The subsidiary condition
\begin{equation}
\sum_\nu \left[u_\nu(\vett r,t)u^*_\nu(\vett r',t)-v_\nu(\vett r',t)v^*_\nu(\vett r,t)\right]=\delta(\vett r-\vett r')
\end{equation}
embodies the commutation rules on the field operators.

In a
uniform fluid the state index $\nu$ is the wave vector $\vett
k$. Equations (3) - 
(5) are solved by taking $u_{\vett k}(\vett r,t)=u_{\vett k}
\exp[i(\vett k\cdot \vett r-\varepsilon_{\vett k}t)]$ and $v_{\vett k}(\vett r,t)=v_{\vett k}
\exp[i(\vett k\cdot \vett r-\varepsilon_{\vett k}t)]$, with the results
\begin{equation}
u_{\vett k}^2=\frac{1}{2}\left\{1+\varepsilon_{\vett k}^{-1}\left[n_0V(k)+\frac{k^2}{2m}\right] \right\}
\end{equation}
and
\begin{equation}
v_{\vett k}^2=-\frac{1}{2}\left\{1-\varepsilon_{\vett k}^{-1}\left[n_0V(k)+\frac{k^2}{2m}\right] \right\}
\end{equation}
where
\begin{equation}
\varepsilon_{\vett
k}=\left[\frac{n_0k^2V(k)}{m}+\left(\frac{k^2}{2m}\right)^2\right]^{1/2}\;.
\end{equation}
The condensate density is given by
\begin{equation}
n_0=n-\langle\tilde\psi^\dagger(\vett r)\tilde \psi (\vett
r)\rangle=n-\sum_{\vett k\neq 0}v_{\vett k}^2
\end{equation}
and the one-body density matrix is
\begin{equation}
\rho(r)=n_0+\langle\tilde\psi^\dagger(\vett r)\tilde \psi
(0)\rangle=n_0+\sum_{\vett k \neq 0}v_{\vett k}^2 \exp(i\vett k \cdot
\vett r)\;.
\end{equation}
We have assumed unitary volume in writing Eqs. (9) and (10).

After these transformations and approximations the Hamiltonian (1) has
been reduced to that of a system of independent bosonic excitations
described by the operators $a_{\vett k}$ and $a_{\vett
k}^\dagger$. The extension of the theory to 
finite temperature is then effected by means of the Bose-Einstein
distribution function
\begin{equation}
f_{\vett k}=\left[\exp(\beta \varepsilon_{\vett k})-1\right]^{-1}\;.
\end{equation}
In particular, Eqs (9) and (10) become
\begin{equation}
n_0=n-\sum_{\vett k\neq 0}\left[v_{\vett k}^2+f_{\vett k}(u_{\vett
k}^2+v_{\vett k}^2) \right]
\end{equation}
and
\begin{equation}
\rho(r)=n_0+\sum_{\vett k \neq 0}\left[v_{\vett k}^2+f_{\vett k}(u_{\vett
k}^2+v_{\vett k}^2) \right]\exp(i\vett k \cdot
\vett r)\;,
\end{equation}
respectively.

\section{The three-dimensional charged-boson fluid}
We introduce the parameter $A=(3n_0/nr_s^3)^{1/2}e^2/(a_Bk_BT)$ and
the reduced distance $R=(12 n_0/nr_s^3)^{1/4}(r/a_B)$. Then
Eqs. (9), (12) and (13) can be rewritten as follows:
\begin{equation}
\left[\left(1-\frac{n_0}{n}\right)\left(\frac{n_0}{n}\right)^{-3/4}
\right]_{T=0}= \frac{2^{1/2}r_s^{3/4}}{3^{1/4}\pi}\int_0^\infty
dx\left[\frac{f(x)}{g(x)} -2x^2\right]\;,
\end{equation}
\begin{equation}
\left[\left(1-\frac{n_0}{n}\right)\left(\frac{n_0}{n}\right)^{-3/4}
\right]_{T\neq0}= \frac{2^{1/2}r_s^{3/4}}{3^{1/4}\pi}\int_0^\infty
dx\left\{\frac{f(x)}{g(x)}\left[1+\frac{2}{\exp[Ag(x)]-1}\right] -2x^2 \right\}
\end{equation}
and
\begin{equation}
\frac{\rho(r)}{n}=1-
 \frac{2^{1/2}}{3^{1/4}\pi}\left(\frac{n_0r_s}{n}\right)^{3/4}\int_0^\infty
dx \left[1-\frac{\sin(Rx)}{Rx}\right]\left\{\frac{f(x)}{g(x)}\left[1+\frac{2}{\exp[Ag(x)]-1}\right]
 -2x^2 \right\} 
\;,
\end{equation}
where we have defined $f(x)=1+2x^4$ and $g(x)=(1+x^4)^{1/2}$. Of
course, Eq.~(14) can also be obtained from Eq.~(15) in the 
limit $T\rightarrow 0$. In the following we use the energy $e^2/a_B$
as the unit of the 
thermal energy $k_BT$.

A numerical solution of Eqs.~(14) and (15) for $n_0/n$ in the physical
range $0\leq n_0/n\leq 1$ can be found for all values of the system
parameters.  We
should bear in mind, however, that we are using a weak-coupling theory
so that the results can be significant only when the depletion of the
condensate is small ({\it i.e.} for $n_0/n$ 
 close to unity, in line of principle). This statement is
quantitatively tested in Table~\ref{tab1}, where we compare our results for $n_0/n$
at $T = 0$ with the DMC data of Moroni {\it et al.}\cite{venticinque} over the whole fluid
range up to Wigner crystallization. It is clear from Table~\ref{tab1} that,
rather surprisingly, the Bogoliubov approach is almost fully
quantitative up to $r_s\simeq 5$, {\it i.e.} for values of $n_0/n$
down to almost 0.5. Very similar results are obtained in the same
range of $r_s$ by the integro-differential equations approach of
Cherny and Shanenko\cite{cherny}.

Table~\ref{tab2} reports our results for $n_0/n$ as a function of temperature
for $r_s=1$   and $r_s= 2$. In the lack of data for a quantitative
test, one may 
hope from the test shown in Table~\ref{tab1} that the Bogoliubov approach could
again be reasonably accurate for values of $n_0/n$ larger than
0.5. According to this crude criterion, it appears from Table~\ref{tab2} that
at such weak couplings the theory could perhaps be useful up to a
fairly sizable value of the reduced temperature $k_BTa_B/e^2$ -
perhaps as large 
as unity for $r_s\simeq 1$.      

Finally, the one-body density matrix $\rho(r)$ (in units of the particle
density $n$) is shown in Figure 1 for $ r_s= 1$ (left panel) and $r_s = 2$
(right panel), at various values of the reduced temperature. The
asymptotic value of $\rho(r)/n$ in the limit $r\rightarrow \infty$ is
the condensate fraction $n_0/n$, 
that we have already presented in Table~\ref{tab2}.

\section{The quasi-two-dimensional charged-boson fluid}
Taking $V(k)=2\pi e^2/k$, Eq. (9) for the condensate fraction at $T =
0$ yields
\begin{equation}
\left[\left(1-\frac{n_0}{n}\right)\left(\frac{n_0}{n}\right)^{-2/3}
\right]_{T=0}= r_s^{2/3}\int_0^\infty
dx\left[\frac{f(x)}{g(x)} -2x^3\right]\;,
\end{equation}
where we have defined $f(x)=1+2x^6$ and $g(x)=(1+x^6)^{1/2}$. The
numerical solution of Eq. (17) 
yields the values of the condensate fraction which are reported in
Table~\ref{tab3} over a range of values for the coupling strength well
below the Wigner phase transition\cite{tanatar} at $r_s\simeq 35$.
We should
expect that the role of correlations becomes more important in lowered
dimensionality, and indeed the crude criterion introduced in Sect.~3
suggests that in the Q2D fluid at $T = 0$ the Bogoliubov approach may be
useful only up to $r_s \simeq 1$.

It is easily seen that the condensate fraction must instead vanish at
$T\neq  0$. Indeed, if we assume $n_0/n\neq 0$ then Eq. (12) yields
\begin{equation}
\left[\left(1-\frac{n_0}{n}\right)\left(\frac{n_0}{n}\right)^{-2/3}
\right]_{T\neq0}= r_s^{2/3}\int_0^\infty
dx\left\{\frac{f(x)}{g(x)}\left[1+\frac{2}{\exp[A x g(x)]-1}\right] -2x^3 \right\}
\end{equation}
where we have set $A=2(n_0/nr_s^2)^{2/3}e^2/(a_Bk_BT)$. The second
term in the square bracket on 
the RHS of Eq. (18) yields a contribution of order $x^{-1}$ to the integrand
for $x\rightarrow 0$, so that the integral diverges. On the other
hand, the solution 
$ n_0/n= 0$ at $T \neq 0$ is consistent with the Bogoliubov - de Gennes
equations: in this case they yield $ v_{\vett k}= 0$ and $u_{\vett k}= 1$, and
the particle  
density is related to the (now finite) chemical potential by
\begin{equation}
n=\sum_{\vett k \neq 0}\left\{ \exp\left[\beta
\left(\frac{k^2}{2m}-\mu\right)\right]-1\right\}^{-1}\;. 
\end{equation}

We can at this point introduce the quasicondensate for the Q2D-CBF at
finite temperature. The essential point is the power-law decay of the
one-body density matrix, which becomes  slow at sufficiently low
temperature and weak
coupling\cite{trenta,trentuno,trentadue,trentatre}. We first show this
by an 
heuristic procedure that we shall justify later below. We isolate the
singular term in $\rho(r)$ and resum it to infinite order to obtain
\begin{equation}
\rho(r)=\tilde \rho(r)\exp[-\Lambda(r)]\;,
\end{equation}
where with the help of Eq.~(13) we have set
\begin{eqnarray}
\Lambda(r)&=&\int \frac{d^2k}{(2\pi)^2}[1-\cos(\vett k\cdot \vett
r)] \frac{V(k)}{\tilde \varepsilon_{\vett k}}f(\tilde \varepsilon_{\vett k})
\nonumber \\&=&2 \left(\frac{n r_s^2}{\tilde n_0}\right)^{1/3}\int_0^\infty
dx\,\frac{1-J_0(x^2R)}{g(x)\left\{\exp[\tilde Ax g(x)]-1 \right\}} 
\end{eqnarray}
and
\begin{eqnarray}
\frac{\tilde \rho(r)}{n}&=&
1-\frac{1}{n}\int\frac{d^2k}{(2\pi)^2}[1-\cos(\vett k\cdot \vett
r)]\left[\frac{1}{2}\left(\frac{\tilde n_0 V(k)+k^2/2m}{\tilde
\varepsilon_{\vett k}}-1\right)+\frac{k^2/2m}{\tilde \varepsilon_{\vett k}}f(\tilde \varepsilon_{\vett k})\right]
\nonumber \\&=&1-\left(\frac{\tilde
    n_0r_s}{n}\right)^{2/3}\int_0^\infty dx\,[1-J_0(x^2R)]\left\{\frac{f(x)}{g(x)}+\frac{4x^6}{g(x)\left(\exp[\tilde Ax g(x)]-1\right)}-2x^3 \right\}\;,
\end{eqnarray}
the quasicondensate density $\tilde n_0$ being
defined by
\begin{equation}
\tilde n_0=\lim_{r\rightarrow \infty}\tilde \rho(r)\;.
\end{equation}
In these equations $\tilde \varepsilon_{\vett k}=(\tilde n_0 V(k)k^2/m+k^4/4m^2)^{1/2}$, $\tilde A=2(\tilde n_0/n r_s^2)^{2/3}e^2/(a_B k_BT)$,
$R=2(\tilde n_0/n r_s^2)^{1/3}(r/a_B)$ and $J_0(y)$ is the
Bessel function of zero-th order.

Figure 2 (upper panels) reports our numerical results for $\rho(r)$ at
two values of the 
coupling strength and at various values of the reduced temperature. It
is evident that at low temperature the decay of the density matrix in
space is  slow and the notion of a quasicondensate thereby
acquires  physical significance. 

A power-law decay of the 
one-body density matrix at low temperature and coupling strength can
be demonstrated directly from Eqs.~(20) - (23).  The function $J_0(x^2R)$ in
the integrand in Eq. (21) provides a lower limit of integration
going as $r^{-1/2}$,
while the upper limit is set by a cut-off wavevector $k_0=1/L$
associated with the 
quasicondensate region (see Popov\cite{trentadue}). 
Eq.~(21) thus yields
\begin{equation}
\Lambda(r)\rightarrow 2 \tilde A^{-1}(n r_s^2/\tilde
n_0)^{1/3}\int_{r^{-1/2}}^{L^{-1/2}}dx\,\frac{1}{x}= 2 \tilde A^{-1}(n r_s^2/\tilde
n_0)^{1/3}\ln(r/L)^{1/2} \;.
\end{equation}
Hence, since $\tilde \rho(r)\rightarrow \tilde n_0$ we find from Eq.~(20)
\begin{equation}
\rho(r)\rightarrow \tilde n_0 (r/L)^{-\alpha}
\end{equation}
with the value of the exponent given by
\begin{equation}
\alpha=\frac{n}{2\tilde n_0}r_s^2 \frac{k_BT}{e^2/a_B}\;.
\end{equation}
Figure~\ref{fig2} (bottom panels) evidentiates the power-law decay of
the density matrix by plotting the curves in log-log scale, as
compared with the aympotic behaviour predicted by Eqs.~(25) and (26) (dots in
the Figure).

\subsection{Asymptotic behaviour of the single-particle Green's function}

We can actually show that
the power-law decay of $\rho(r)$ that we have obtained just above derives
from the correlations between phase fluctuations in the Q2D-CBF. We
follow the method proposed for the neutral 2D gas in the  work of
Popov\cite{trentadue} (see also Fisher and
Hohenberg\cite{trentatre}). The  
fluctuations of the phase $\phi(\vett x,t)$ determine the single-particle
Green's 
function in the low-momentum regime (below the cut-off momentum $k_0$)
according to
\begin{equation}
G(\vett x,\tau;\vett x_1,\tau_1)\simeq \tilde
n_0\exp\left\{-\frac{1}{2}\langle\left[ \phi(\vett x,\tau)-\phi(\vett
    x_1,\tau_1)\right]^2 \rangle \right\}\;. 
\end{equation}
From Eq.~(19.16) in Chapter 6 of Popov's book\cite{trentadue} we find
\begin{equation}
\langle\phi(\vett k,\omega) \phi(-\vett k,-\omega)
\rangle\rightarrow\frac{V(k)}{\omega^2+\tilde n_0V(k)k^2/m}
\end{equation}
for the Q2D-CBF at long wavelengths and frequencies, so that
\begin{eqnarray}
\frac{1}{2}\langle\left[ \phi(\vett x,\tau)-\phi(\vett
    x_1,\tau_1)\right]^2
    \rangle\rightarrow\frac{1}{2\beta}\sum_{k<k_0}\sum_{\omega} 
\frac{V(k)}{\omega^2+\tilde n_0V(k)k^2/m}\nonumber \\\times |\exp[i(\vett k
    \cdot \vett x-\omega \tau)]-\exp[i(\vett k
    \cdot \vett x_1-\omega \tau_1)]|^2\;.
\end{eqnarray}
In Eq.~(29) we carry out the summation over the Matsubara
frequencies for $\tau_1=\tau^+$  and keep the most diverging contribution in the integral
over the momenta to
obtain 
\begin{equation}
\frac{1}{2}\langle\left[ \phi(\vett x,\tau)-\phi(\vett
    x_1,\tau^+)\right]^2
    \rangle\rightarrow\alpha \ln\frac{r}{L}+ {\rm const}.
\end{equation}
where $r=|\vett x-\vett x_1|$ and
the quantity $\alpha$ is given by Eq.~(26). Using this result in Eq.~(27) we
see that the Green's function decays to zero with the law $r^{-\alpha}$.

\subsection{Superfluid fraction and quasicondensate fraction}

As already noted in Sect.~1, the notion of a 
quasicondensate becomes meaningful at temperatures below the
Kosterlitz-Thouless transition. A superfluid component is therefore
present at these temperatures and its density $n_s$ is given by\cite{trentatre}
\begin{equation}
\frac{n_s}{n}=1-\frac{\beta}{2nm}\sum_{\vett k\neq 0}k^2\frac{\exp(\beta
  \varepsilon_{\vett k})}{[\exp(\beta
  \varepsilon_{\vett k})-1]^2}
\end{equation}
From Eq.~(31) we have
\begin{equation}
\frac{n_s}{n}=1-2 \tilde A\left(\frac{r_s \tilde
    n_0}{n}\right)^{2/3}\int_0^\infty dx\,\frac{\exp\left[\tilde A (x+x^4)^{1/2}\right]}{\left\{\exp\left[\tilde A (x+x^4)^{1/2}\right]-1\right\}^2}\;.
\end{equation}
The value of the
superfluid fraction $n_s/n$ that we obtain from Eq.~(32) are reported in
Table~\ref{tab4} for two values of the coupling strength and at various values
of the reduced temperature.

In Table~\ref{tab4} we also show the values taken by the
quasicondensate fraction $\tilde n_0/n$ at the same values of the
coupling strength. The quasicondensate density is similar to the
superfluid density  at very low coupling, but rapidly decreases as the
coupling strength is increased.

\section{Summary and outlook}
Summarizing, we have studied quasicondensation and superfluidity in a weakly
interacting 2D fluid of charged bosons with $e^2/r$ interactions at finite
temperature. By comparison of results on the 3D fluid with diffusion Monte
Carlo data~\cite{venticinque} we have found that the Bogoliubov approach may
yield quantitatively useful predictions over a surprisingly wide range of
coupling strengths and of deviations of the condensate fraction from unity. At
finite temperature the behaviour of the charged 2D gas is qualitatively wholly
similar to that of its better known neutral analogue: well below the
Kosterlitz-Thouless transition a slow power-law decay is seen in the one-body
density matrix, heralding extended-range correlations in phase fluctuations
and the formation of a condensate over regions of finite size.

From the diffusion Monte Carlo results of Magro and Ceperley \cite{ventisei}
it appears that the charged 2D fluid with logarithmic interactions may also
show quasicondensation, even in the absence of macroscopic condensation in the
ground state. Other interesting questions in this area regard how the
transition from 3D to 2D behaviour is effected and how one may develop a sound
theoretical description of the momentum distribution with increasing coupling
strength.  

\acknowledgments
This work was partially supported by MURST under the PRIN-2000 Initiative.

\begin{table}
\caption{Condensate fraction in the 3D-CBF at zero temperature from the
  Bogoliubov approach (B), compared with the diffusion Monte Carlo data (DMC,
  from Moroni {\it et al.} {\protect \cite{venticinque}}).}
\begin{center}
\begin{tabular}{ccccccccc}
$r_s$ & 1 & 2& 5 &10 & 20 & 50 & 100 &160\\ \hline
$(n_0/n)_B$& 0.818 & 0.722& 0.549& 0.401& 0.264& 0.132& 0.072&0.047 \\
$(n_0/n)_{DMC}$ & 0.827& 0.722& 0.542& 0.359& 0.206& 0.053& 0.0104&
0.004\\
\end{tabular}
\end{center}
\label{tab1}
\end{table}

\begin{table}
\caption{Condensate fraction in the 3D-CBF as a function of the reduced
  temperature from the Bogoliubov approach, for two values of the coupling
  strength $r_s$.}
\begin{center}
\begin{tabular}{ccccc}
$T$& $(n_0/n)_{r_s=1}$&$\;\;\;\;\;$&$T$&$(n_0/n)_{r_s=2}$\\
\hline
0 & 0.818& &0 &0.722\\
0.3& 0.816& &0.1& 0.718\\
0.5& 0.793& &0.2& 0.652\\
0.75& 0.720& &0.3& 0.466\\
1.0& 0.593& &0.32& 0.410\\
1.25& 0.407& &0.34& 0.345\\
1.4& 0.247& &0.36& 0.266\\
1.482& 0.092& &0.38& 0.153\\
\end{tabular}
\end{center}
\label{tab2}
\end{table}

\begin{table}
\caption{Condensate fraction in the Q2D-CBF at zero temperature from the
  Bogoliubov approach.}
\begin{center}
\begin{tabular}{cccccccccc}
$r_s$& 0.01& 0.1& 0.2& 0.4& 0.6& 0.8& 1& 2& 5\\
\hline
$n_0/n$& 0.968& 0.863& 0.794& 0.700& 0.633& 0.580& 0.537& 0.398& 0.230\\
\end{tabular}
\end{center}
\label{tab3}
\end{table}

\begin{table}
\caption{Superfluid and quasicondensate fraction in the Q2D-CBF as a
  function of reduced 
  temperature from the Bogoliubov approach, for two values of the coupling
  strength $r_s$.}
\begin{center}
\begin{tabular}{ccccccc}
$T$&
$(n_s/n)_{r_s=0.1}$&$(\tilde n_0/n)_{r_s=0.1}$ &$\;\;\;\;\;\;$&$T$&$(n_s/n)_{r_s=1}$&$(\tilde n_0/n)_{r_s=1}$\\ 
\hline
0 &1&0.863 && 0&1&0.537\\
10& 0.991 & 0.861 && 0.2& 0.995 &0.536\\
15& 0.972 & 0.854 && 0.3& 0.979 &0.533\\
20& 0.944 & 0.843 && 0.5& 0.905 &0.517\\
30& 0.865 & 0.809 && 0.7& 0.779 &0.482\\
40& 0.763 & 0.758 && 0.8& 0.696 &0.457\\
\end{tabular}
\end{center}
\label{tab4}
\end{table}
\noindent
\begin{figure}
\centerline{\psfig{file=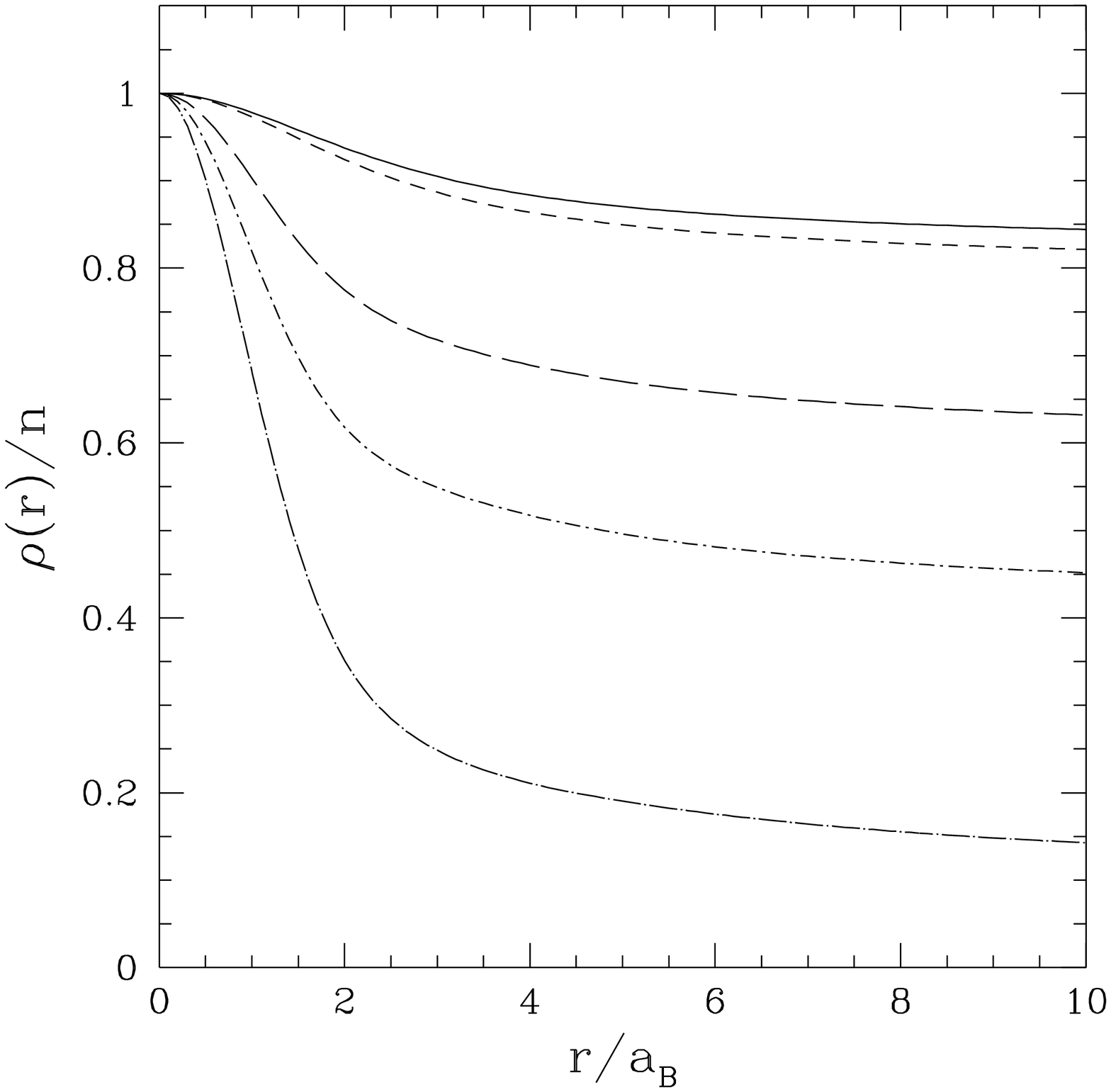,width=0.4\linewidth}\psfig{file=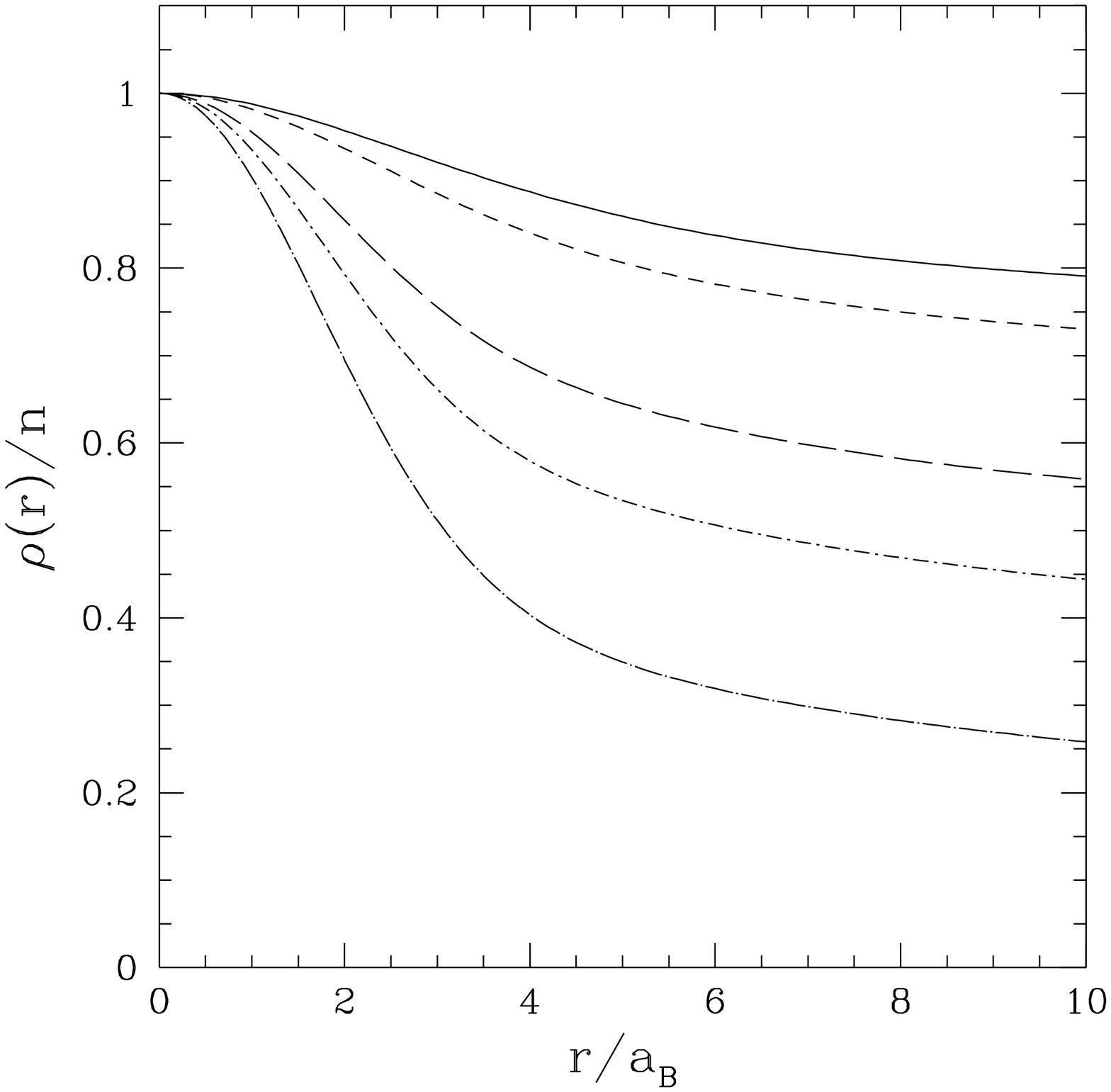,width=0.4\linewidth}}
\caption{The one-body density matrix $\rho(r)/n$ as a function of $r/a_B$ in
  the 3D-CBF. Left: for $r_s$=1 at values of the reduced temperature $T$ equal
  to 0, 0.5, 1.0, 1.25, and 1.482 (from top to bottom). Right: for $r_s=2$ at
  $T=$0, 0.2, 0.3, 0.34 and 0.38.}
\end{figure}
\newpage
\begin{figure}
\centerline{\psfig{file=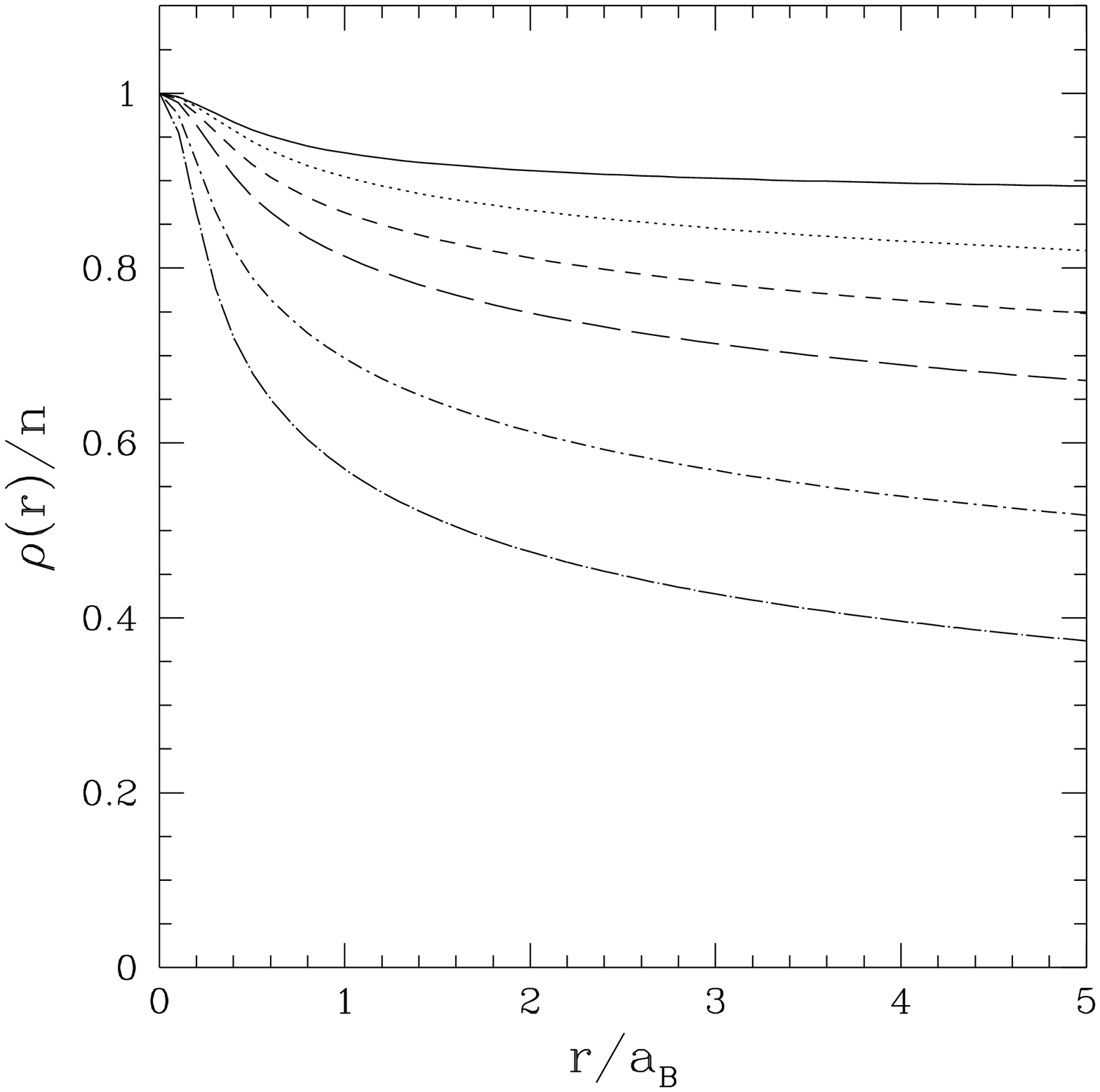,width=0.4\linewidth}\psfig{file=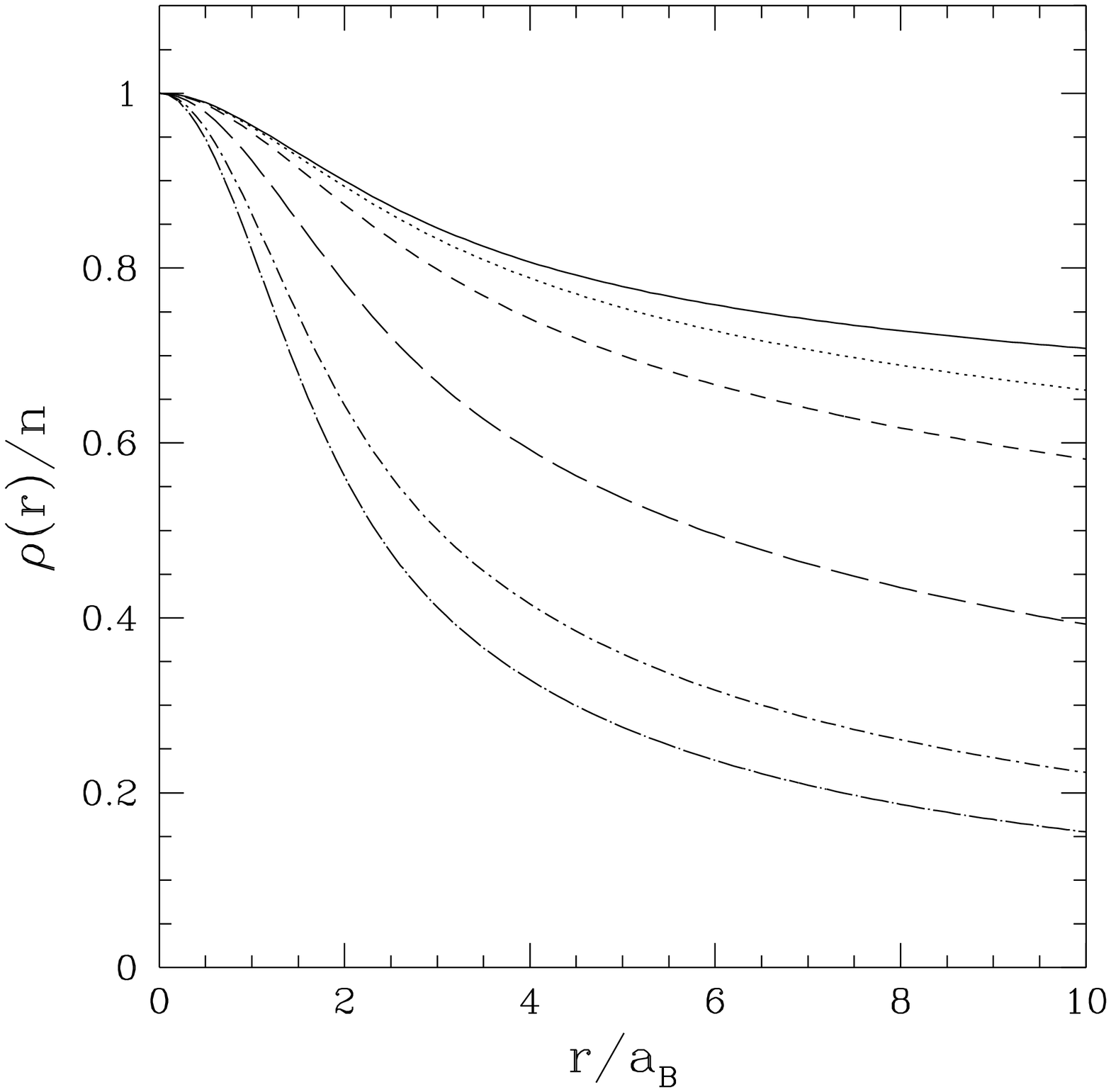,width=0.4\linewidth}}
\centerline{\psfig{file=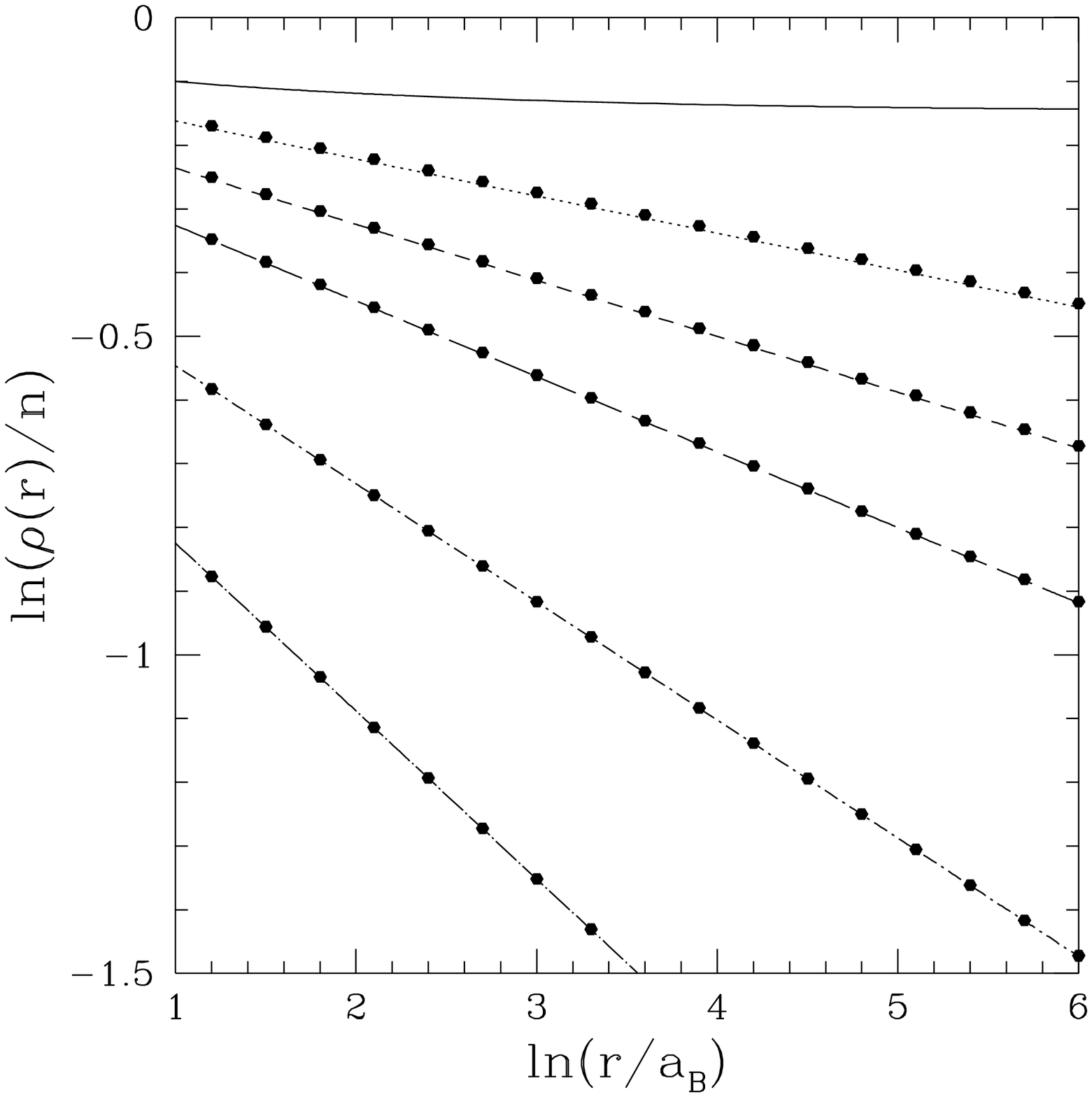,width=0.4\linewidth}\psfig{file=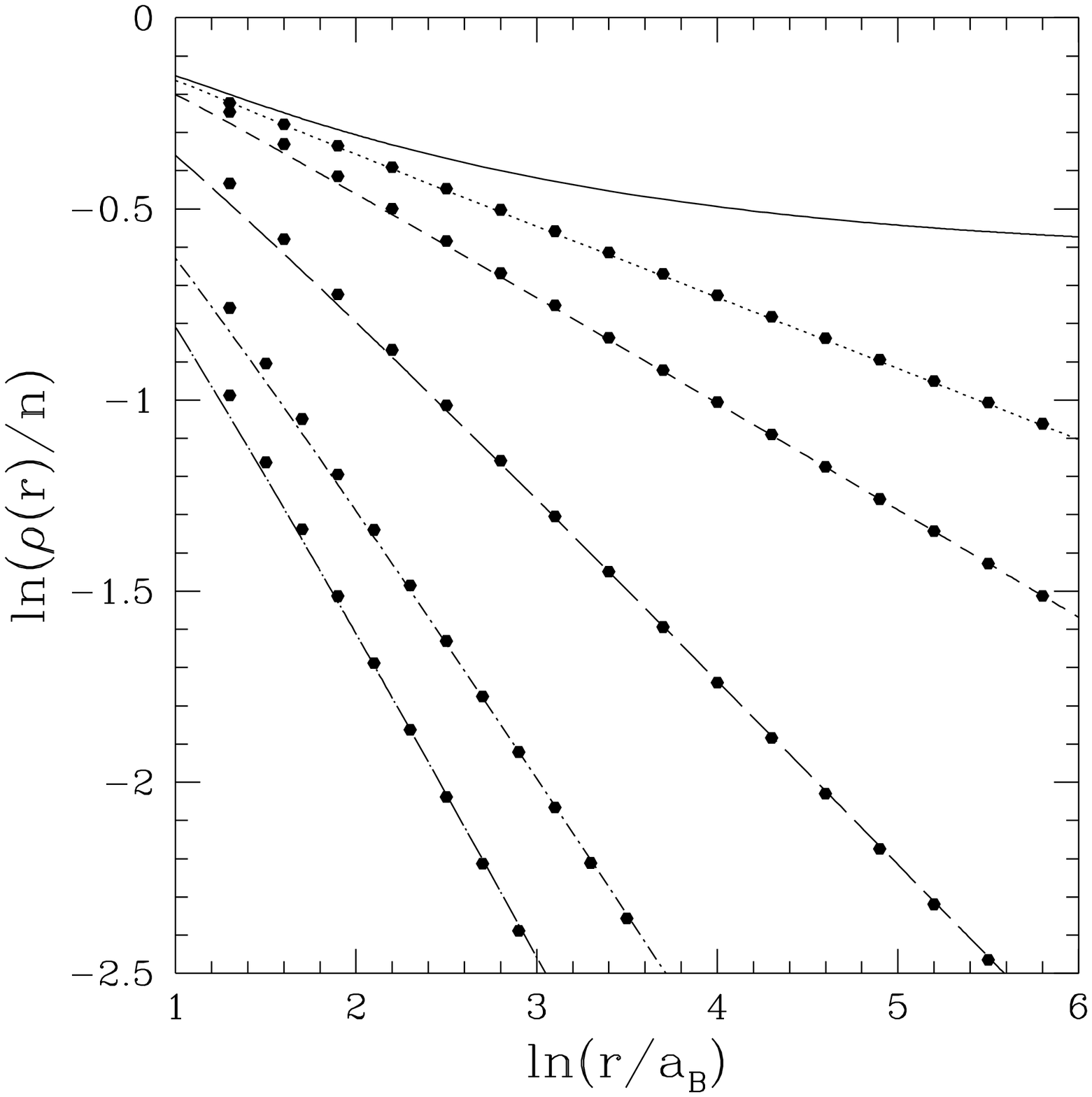,width=0.4\linewidth}}
\caption{The one-body density matrix $\rho(r)/n$ as a function of $r/a_B$ in
  the Q2D-CBF, both in linear scale (upper panels) and in logarithmic scale
  (lower panels). Left panels: for $r_s$=0.1 at values of the reduced
  temperature $T$ 
  equal to 0, 10, 15, 20, 30 and 40 (from top to bottom). Right panels
  : for $r_s=1$
  at $T=0$, 0.2, 0.3, 0.5, 0.7 and 0.8. The dots in the lower
  panels give the asympotic behaviour of $\rho(r)/n$ 
  at large $r$ as evaluated
  analytically  in
  Eqs.~(25) and (26). The length $L$ in Eq.~(25), which is not determined by
  the asymptotic 
  calculation, appears in logarithmic scale as an additive constant 
  and is here fixed  by requiring that the analytic result overlaps
  the numerical one as $r$ increases.}
\label{fig2}
\end{figure}

\end{document}